\documentclass{PoS}
\usepackage{wrapfig}
\usepackage{caption}
\usepackage{subcaption}

\usepackage{lineno}

\usepackage{graphicx}
\usepackage{amsmath,amsfonts,amssymb}
\usepackage{nicefrac}
\usepackage{graphicx,wrapfig,lipsum}
\usepackage{lineno}

\title{Comparison of the measured atmospheric muon flux with Monte Carlo simulations for the first KM3NeT detection units}

\ShortTitle{Measured and simulated atmospheric muons for the first KM3NeT detectors}

\author{
The KM3NeT Collaboration$^{\ddagger}$\footnote{for collaboration list see PoS(ICRC2019)1177}\\
{$^{\ddagger}$ \itshape \href{https://www.km3net.org/km3net-author-list-for-icrc-2019}{https://www.km3net.org/km3net-author-list-for-icrc-2019}}\\
E-mail: \email{piotr.kalaczynski@ncbj.gov.pl}
}

\abstract{
The KM3NeT Collaboration has successfully deployed its first detection units in the Mediterranean Sea in December 2015 (ARCA) and September 2017 (ORCA). The sample of data collected between September 2016 and March 2017 has been used to measure the atmospheric muon flux at two different depths under the sea level: about 3.5\,km with ARCA and about 2.5\,km with ORCA. The atmospheric muon flux represents an abundant signal for a neutrino telescope and can be used to test the reliability of the Monte Carlo simulation chain. In this work, the measurements are compared to Monte Carlo simulations based on MUPAGE and CORSIKA codes. Measured events and simulated events are treated using the same approach, making the comparison reliable.
\\

\vspace{4mm}
{\bfseries Corresponding authors:}
\speaker{{Piotr Kalaczynski}}$^{1}$,
Rosa Coniglione$^{2}$
\\
{$^{1}$ \itshape National Centre for Nuclear Research, 00-681 Warsaw, Poland}\\
{$^{2}$ \itshape Instituto Nazionale Di Fisica Nucleare - Laboratori Nazionali del Sud, Via S. Sofia 62 - 95123 Catania, Italy
}
}

\FullConference{36th International Cosmic Ray Conference 2019\\
                25.07.2019 - 1.08.2019\\
                Madison, Wisconsin, USA}

\begin{document}

\section{Introduction}

KM3NeT is a research infrastructure being constructed at the bottom of the Mediterranean Sea. It consists of two neutrino detectors: ARCA (Astroparticle Research with Cosmics in the Abyss) located off-shore Portopalo di Capo Passero, Sicily, Italy, at a depth of 3500\,m and ORCA (Oscillation Research with Cosmics in the Abyss) off-shore Toulon, France, at a depth of 2450\,m.

The main purpose of the ARCA telescope is the detection of TeV-PeV neutrinos from astrophysical sources or in coincidence with other high energy events, for example gravitational waves, gamma ray bursts, blazar flares. The ORCA telescope is designed to study the neutrino mass hierarchy (NMH), using oscillations of the atmospheric neutrinos in the GeV range \cite{LoI}.

Both KM3NeT detectors are based on the same technology \cite{LoI}: vertically aligned detection units (DUs), each carrying 18 digital optical modules (DOMs). Each DOM contains 31 3-inch photomultiplier tubes (PMTs), calibration and positioning instrumentation and readout electronics boards. The difference between ARCA and ORCA is the horizontal (90\,m and 20\,m respectively) and vertical (36\,m and 9\,m respectively) spacing between the DOMs, which is optimised to search for high- and low-energy neutrinos, respectively. The first DUs are already taking data at both the ARCA and ORCA sites. Atmospheric muons represent an important background to the physics analyses in a neutrino telescope, but they are also a useful tool to test the performance of the detector and validity of the simulations. In this paper a comparison of the first data collected so far by ARCA and ORCA to the Monte Carlo (MC) simulations is presented.

\section{Simulation chain}

\begin{figure}
    \centering
    \includegraphics[width=0.8\textwidth]{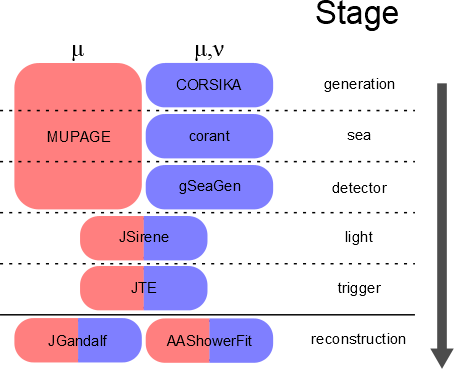}
    \caption{Schematic drawing of the KM3NeT atmospheric muon simulation chain. Programs are colour-coded, corresponding to the physics event generator for which they are used. The MUPAGE chain is marked in red and the CORSIKA chain in blue. The software is grouped according to the simulation stage it belongs to.
    \label{fig:SimChain}}
\end{figure}%

The KM3NeT simulation chain shown in Figure \ref{fig:SimChain} has a modular structure. The generation of the atmospheric muon bundles is performed using two different codes: MUPAGE \cite{MUPAGE} and CORSIKA \cite{CORSIKA}. MUPAGE is a fast, parametric simulation code developed for the ANTARES experiment \cite{ANTARES} that generates muon bundles, induced by cosmic rays impinging the Earth atmosphere, at different undersea depths and different zenith angles, on the surface of the active volume of the detector. CORSIKA (COsmic Ray SImulations for KAscade) is a software package that simulates the interactions of the primary cosmic ray (CR) nuclei in  the upper layers of the atmosphere and follows the development of the shower to a specified observation level (in this work it is the sea level). Their energy spectrum  is simulated following a power law, \(E^{-\gamma}\), where the spectral index \(\gamma\) is chosen by the user. The CORSIKA package is flexible and allows the user to choose between different models to describe the primary  interactions and the composition of the CR primary flux.

A dedicated software, called "Corant", has been developed to convert the CORSIKA output format to the KM3NeT standard format of events and to evaluate the weights to be applied to each event, according to the model of the CR composition chosen by the user. The propagation of the muon bundles to the active volume of the detector is performed using a 3-dimensional muon propagator contained in the gSeaGen \cite{gSeaGen} library, a GENIE-based \cite{GENIE} package designed mainly to simulate neutrino interactions. After the muon bundle propagation, the Cherenkov photons emitted along the path of muons and the probability of their detection by the DOMs are simulated with a custom application developed for KM3NeT using multidimensional interpolation tables \cite{JSirene}, called "JSirene". The next step is the creation of a simulated data stream, similar to the data stream coming from the detector. This is done with  the addition of the environmental optical background, due to the bioluminescence and to the \(^{40}\mathsf{K}\) decay, with the simulation of the detector response, taking into account the front-end electronics behaviour. The same trigger algorithms used for real data are used to identify possible interesting events in the simulated sample \cite{AtmMuRate}. Finally, the simulated events are processed with the same reconstruction programs used for the real data stream. At this stage the simulated MC events can be easily compared to the calibrated data. A  track reconstruction algorithm is applied both to data and MC events. In this work only a track reconstruction code ("JGandalf" \cite{reco}) is considered, which is specifically designed to reconstruct atmospheric muons and muons induced by neutrinos.

\section{Data vs MC comparisons}

\begin{figure}
     \centering
     \includegraphics[width=0.8\textwidth]{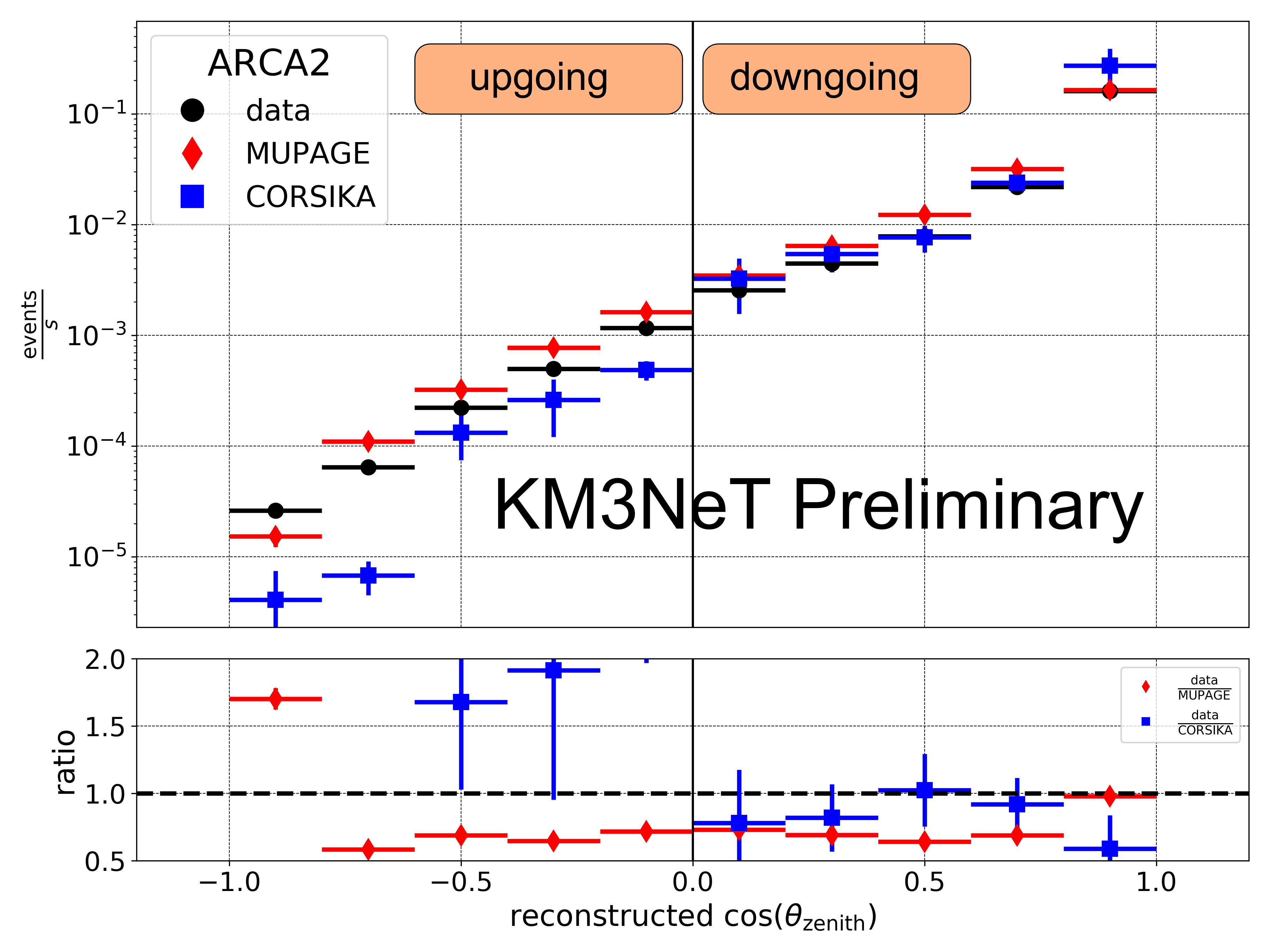}
     \caption{Rate of atmospheric muons as a function of the reconstructed zenith angle for data and MC simulation for the ARCA2 detector. Most of the upgoing events (all MC events) are downgoing events that are badly reconstructed as upgoing. No quality cuts have been applied to remove the badly reconstructed tracks.}
     \label{fig:zenith-1}
\end{figure}

\begin{figure}
     \centering
     \includegraphics[width=0.8\textwidth]{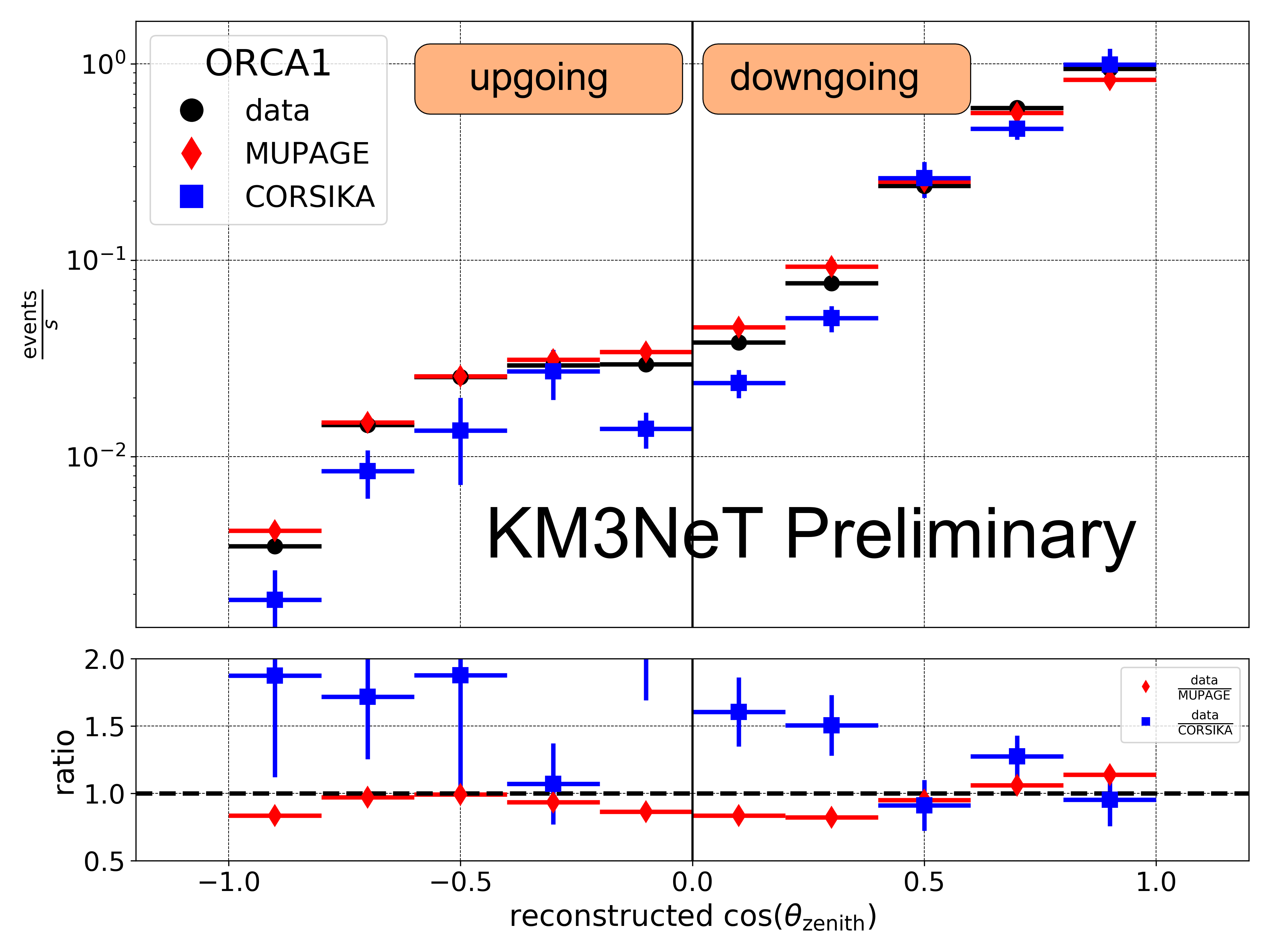}
     \caption{Rate of atmospheric muons as a function of the reconstructed zenith angle for data and MC simulation for the ORCA1 detector. Most of the upgoing events (all MC events) are downgoing events that are badly reconstructed as upgoing. No quality cuts have been applied to remove the badly reconstructed tracks.}
     \label{fig:zenith-2}
\end{figure}

\begin{figure}
    \centering
    \includegraphics[width=0.8\textwidth]{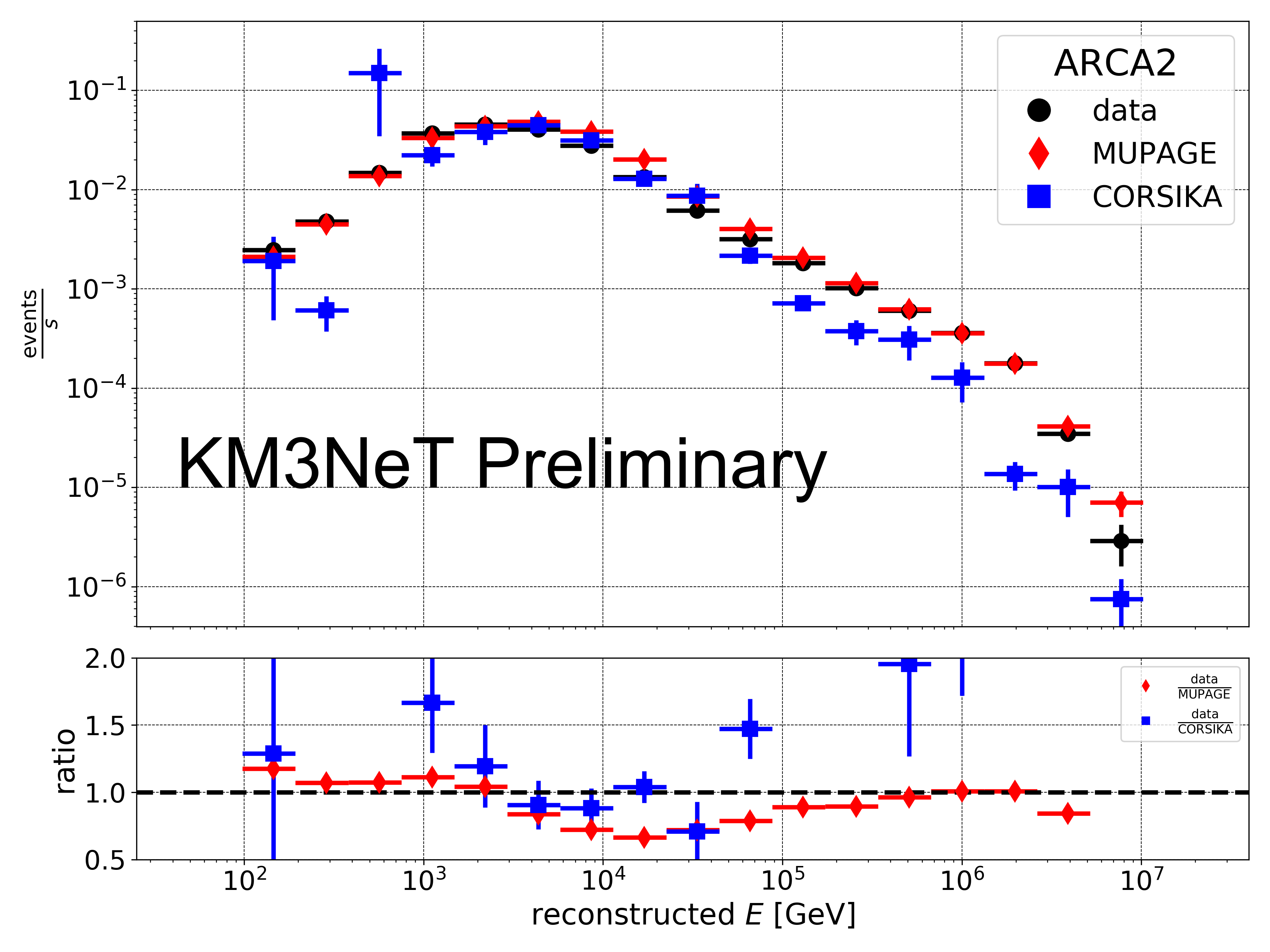}
    \caption{Rate of atmospheric muons as a function of the reconstructed energy for data and MC simulation  for the ARCA2 detector. More details on the energy reconstruction can be found in \cite{reco}.}
    \label{fig:energy-1}
\end{figure}%

A sample of data collected with 2 DUs of the ARCA detector (ARCA2) from the 23th of December 2016 to the 2nd of March 2017 and with 1 DU of the ORCA detector  (ORCA1) from the 28th of September 2017 to the 13th of December 2017 has been reconstructed with the track reconstruction algorithm "JGandalf" (see Figure \ref{fig:SimChain}). 

A sample of muon bundles with energy larger than 10 GeV and multiplicity up to 100 tracks was simulated with MUPAGE. The equivalent livetime is close to  the considered detector livetime, about 20\,days for ARCA2 and 23\,days for ORCA1. The muon bundle energy is evaluated at the surface of the detector sensitive volume.
The events processed as described in Section 2 are shown in Figures \ref{fig:zenith-1}, \ref{fig:zenith-2} and \ref{fig:energy-1} with the red marks. Only statistical errors are shown in the plots.  

A total number of \(2.5\cdot10^9\) showers have been simulated with CORSIKA, using SIBYLL-2.3c to describe the high-energy hadronic interaction model \cite{SIBYLL}. Five different species of nuclei were considered: \(p\), \(He\), \(C\), \(O\) and \(Fe\) with energies between \(10^3\) and \(10^9\)\,GeV and an energy spectrum \(E^{-1}_{\mathsf{primary}}\).
For the muon bundles generated with CORSIKA, the simulation procedure presented in Section 2 has been followed and events were reconstructed with the same set of programs as for MUPAGE and data events. At the end of the reconstruction, the events have been weighted according to the composition model described in \cite{GST}.
The results of the CORSIKA simulation are shown as blue markers in Figures \ref{fig:zenith-1}, \ref{fig:zenith-2} and \ref{fig:energy-1}. The errors shown are calculated as \(\Delta x=\sqrt{{\sum}w_{i}^{2}}\), where \(w_{i}\) are the weights of each event. 
In all plots no systematic uncertainties on the simulation are considered.

The preliminary results discussed in this paper show that the considered  MC samples can reproduce the data with a good level of agreement. The upgoing muon contribution in the zenith angle plots (Figures \ref{fig:zenith-1} and \ref{fig:zenith-2}) is due to badly reconstructed downgoing muons. In this work, no quality cuts or selections have been applied to remove the poorly reconstructed tracks. Notice that expected systematic uncertainties, which have not been considered here, are much larger than the statistical errors. 
The good data/MC agreement is not surprising as both approaches, the parameterized simulation with MUPAGE and the full simulation with CORSIKA, have been successfully used in ANTARES \cite{ANTARES_MU}, and confirm that the 
simulation chain set up for KM3NeT analyses is reliable and under control.

\section{Conclusions}

Using the data taken with the first KM3NeT detection units, it is possible to compare the measured and the expected muon rate at two different sites. The good level of agreement confirms the reliability of the simulation procedure defined so far within KM3NeT.
Sharing the same technology and detection medium (sea water), the ARCA and ORCA detectors offer a unique opportunity to probe the depth dependence of the atmospheric muon flux without applying systematic uncertainties related to comparison between different types of experiments \cite{AtmMuRate}.

Once the simulation chain is well established, next steps foresee the increase of the simulated samples, particularly at very high energies, and the study of the impact of the physics parameteres. This includes the description of the hadronic interaction models in some regions of the phase space that cannot be explored with accelerators, of the chemical composition of the primary CRs and of the optical properties of deep-sea water. A detailed study of the detection and reconstruction efficiency can be also performed.

\section*{Acknowledgements}
This work was supported by the National Centre for Science: Poland grant 2015/18/E/ST2/00758.
\end{document}